# PAPR reduction of space-time and space-frequency coded OFDM systems using active constellation extension

Mahmoud Ferdosizadeh Naeiny, Farokh Marvasti

Electrical Engineering Department, Sharif University of Technology

mferdosi@ee.sharif.edu, fmarvasti@sharif.edu

October 30, 2018

## Abstract

Active Constellation Extension (ACE) is one of techniques introduced for Peak to Average Power Ratio (PAPR) reduction for OFDM systems. In this technique, the constellation points are extended such that the PAPR is minimized but the minimum distance of the constellation points does not decrease. In this paper, an iterative ACE method is extended to spatially encoded OFDM systems. The proposed methods are such that the PAPR is reduced simultaneously at all antennas, while the spatial encoding relationships still hold. It will be shown that the original ACE method can be employed before Space Time Block Coding (STBC). But in case of Space Frequency Block

Coding (SFBC), two modified techniques have been proposed. In the first method, the OFDM frame is separated by several subframes and the ACE method is applied to these subframes independently to reduce their corresponding PAPRs. Then the low PAPR subframes are recombined based on SFBC relationships to yield the transmitted signals from different antennas. In the second method, for each iteration, the ACE is applied to the antenna with the maximum PAPR, and the signals of the other antennas are generated from that of this antenna. Simulation results show that both algorithms converge, but the second method outperforms the first one when the number of antennas is increased.

Keywords:ACE, OFDM, PAPR, POCS, STBC, SFBC

# 1 Introduction

Spatial diversity systems increase the reliability of wireless networks using several transmitter and receiver antennas simultaneously. To achieve the full diversity at the transmission side, the space time codes have been introduced in [1, 2]. OFDM is another technique used in the frequency selective channels [3]. By combination of spatial diversity and OFDM techniques, a higher capacity can be achieved over broadband multipath fading wireless channels [4, 5]. One of the drawbacks of OFDM systems is high PAPR which leads to the saturation of the high power amplifier. Thus, a high dynamic range amplifier is needed, which increases the cost of the system. Some tech-niques have been proposed to reduce the PAPR in single antenna OFDM systems [6]-[14]. Some of these methods need Side Information (SI) to be transmitted to the receiver such as Partial Transmit Sequence (PTS) [6, 7] and Selected Mapping (SLM) [8, 9]. Some other methods do not need side information such as clipping and filtering [10, 11], tone reservation [12], block coding [13], and ACE [14].

In the ACE method, the constellation points are extended such that the PAPR is minimized without any increase in Bit Error Rate (BER). To find the proper extension of the symbols, an iterative method which is based on Projection Onto Convex Sets (POCS) optimization algorithm has been proposed in [14]. In this method, at each iteration, the time domain signal is clipped and filtered. The clipping and filtering noise moves the constellation points. The extension of each symbol must be within the allowable regions; otherwise, the point is moved to its original position. These procedures are performed iteratively to achieve the target PAPR.

The main problem in extension of ACE method to the spatially encoded OFDM systems is the simultanous PAPR reduction at all antennas while the spatial encoding relationships still hold. Two types of spatial encodings are discussed; STBC and SFBC. In case of STBC, the time domain signal of antennas are the conjugate or negative conjugate of the OFDM frames before the STBC. Thus, if the ACE algorithm is applied to the OFDM frames and then the STBC encoding is applied, the PAPR is reduced for each antenna. In the SFBC case, the spatial encoding is done between the adjacent subcarriers of OFDM frame [4]. To hold the SFBC structure of the frequency domain signal of different antennas and to reduce their PAPR simultaneously, two modified ACE methods are proposed. It will be shown that the time domain signals of different antennas in SFBC is the combination of several signals and their conjugates. These signals can be derived by separation of the frequency domain vector into several subframes and the application of IFFT to these subframes. If the PAPR of these generating signals is reduced, then the PAPR at all antennas can be reduced. This is the base of the first proposed method for SFBC case. In this method the iterative ACE is applied to the subframes independently and then the signal of antennas are generated by the combination of the resulting low PAPR subframes. In the second algorithm for each iteration, the antenna with the maximum PAPR is selected and the clipping, filtering and movement of the constellation points are applied. Then the signals of the other antennas are produced using SFBC relationships. This procedure is done iteratively to achieve the target PAPR or it will be stopped after a number of iterations.

The rest of this paper is organized as follows: In section 2, the single antenna OFDM systems are modeled and the iterative ACE method for PAPR reduction at these systems is discussed briefly. Then in section 3 the system model of STBC-OFDM systems is introduced and the ACE method is extended. In section 4, the SFBC-OFDM systems are investigated and the ACE optimization problem is introduced. Two proposed methods for SFBC are introduced in subsections 4.1 and 4.2. Section 5 is about the computational complexity of the proposed methods. Section 6 is related to the simulation results and performance evaluation.

## 2 Single antenna OFDM systems and the ACE method

In single antenna OFDM systems, the input bit stream is interleaved and encoded by a channel encoder. Then, the coded bits are mapped onto the complex symbols using digital modulation techniques. The sequential symbols are converted to blocks of $N_c$ complex symbols. $\mathbf{S}_m = [S_m(0), S_m(1), ..., S_m(N_c-1)]^T$ is the $m$th block, where $N_c$ is the number of OFDM subcarriers. Then $N - N_c$ zeros are added ($\frac{N}{N_c}$ is the oversampling ratio) before the IFFT block to yield the oversampled time domain vector $\mathbf{s}_m = [s_m(0), s_m(1), ..., s_m(N-1)]^T$:

$$s_m(n) = \frac{1}{\sqrt{N}} \sum_{k=0}^{N_c-1} S_m(k) e^{j\frac{2\pi nk}{N}}, n = 0, 1, ..., N-1. \quad (1)$$

The PAPR of the OFDM frame is defined as the ratio of the maximum to average power of the time domain samples:

$$PAPR_m = \frac{\max_n\{|s_m(n)|^2\}}{E\{|s_m(n)|^2\}} \quad (2)$$

where $E\{.\}$ is the mathematical expectation. Based on (1), the time domain samples are the sum of $N_c$ independent terms. When $N_c$ is large, based on the central limit theorem, the time domain samples have Gaussian distribution and thus they may have large amplitudes.

To reduce the PAPR by the ACE method [14], the complex symbols $S_m(k)$ are extended such that the PAPR of the time domain signal is reduced while the minimum distance of the constellation points does not decrease. Thus at the receiver side the Bit Error Rate (BER) does not increase. Fig.1 shows the regions in which the symbols $S_m(k)$ can be moved such that the minimum distance of the constellation points does not decrease. If the symbol $S_m(k)$ is extended to the point $S_m(k) + C_m(k)$ and $\mathbf{C}_m = [C_m(0), C_m(1), ..., C_m(N_c-1)]^T$, then the following optimization problem must be solved:

$$\min_{\mathbf{C}_m} \left\{ \max_n \{ s_m(n) + \frac{1}{\sqrt{N}} \sum_{k=0}^{N_c-1} C_m(k) e^{j\frac{2\pi nk}{N}} \} \right\}$$
$$Subject \quad to \quad \|\mathbf{C}_m\|^2 \leq \Delta P \qquad (3)$$

where $\Delta P$ limits the power increase. In [14] an iterative algorithm for finding the suboptimum solution of (3) has been proposed. In this method, which has been shown in Fig.2, the time domain samples $s_m(n)$ are clipped as follows:

$$\bar{s}_m(n) = \begin{cases} s_m(n) & \text{if } |s_m(n)| \leq A \\ \frac{A s_m(n)}{|s_m(n)|} & \text{if } |s_m(n)| > A \end{cases}, \qquad (4)$$

where $A$ is the clipping threshold. The nonlinear clipping operation creates the in-band and out-of-band distortions. The out-of-band components must be removed. Therefore the samples are converted again to the fre-quency domain using FFT operation and $N - N_c$ out-of-band components are removed. Note that the in-band components $S_m(k), k = 0, 1, \cdots, N_c - 1$ have been moved from their initial position in the constellation. The new points denoted by $\bar{S}_m(k)$ must be mapped to the regions shown in Fig.1. To achieve this, the points in acceptable regions are kept unchanged and the other points are mapped to these regions. Then the time domain samples are generated again form the extended symbols and the procedure of clipping, removing the out-of-band components and mapping to the acceptable regions is performed iteratively to achieve the target PAPR.

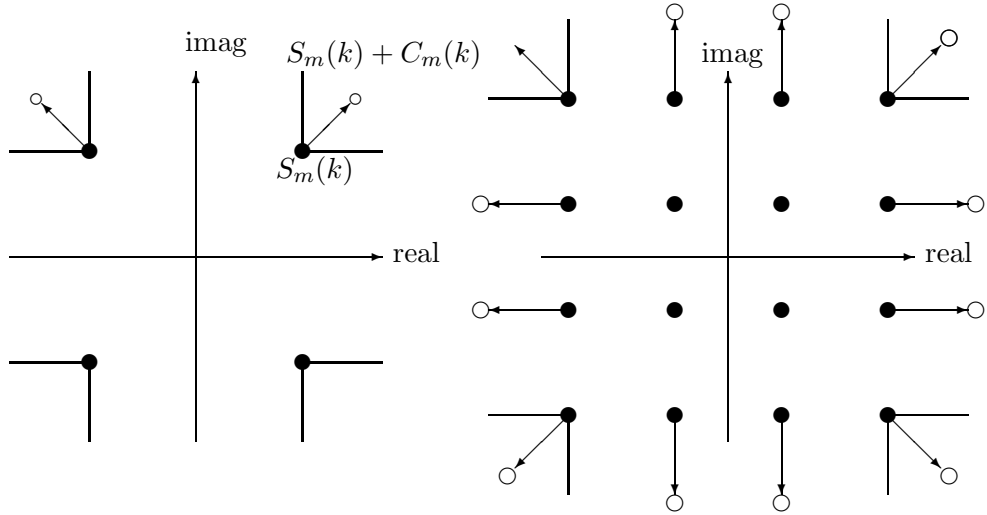

Figure 1: Constellation extension for PAPR reduction of QPSK and 16-QAM modulations

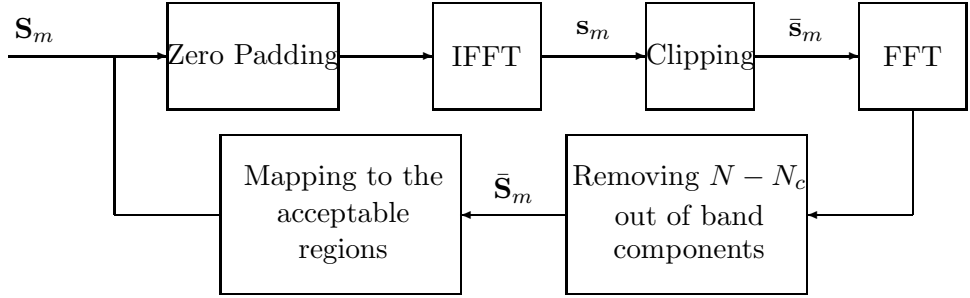

Figure 2: Iterative ACE method for PAPR reduction of single antenna OFDM systems.

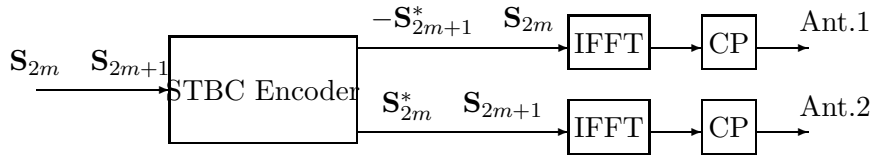

Figure 3: Block diagram of STBC-OFDM system with two transmitter antennas.

# 3 STBC-OFDM systems and ACE method for PAPR reduction

In STBC-OFDM system with two transmitter antennas shown in Fig.3, the space time block coding are applied to two sequential blocks $\mathbf{S}_{2m}$ and $\mathbf{S}_{2m+1}$. If $\mathbf{S}_m^{(p)} = [S_m^{(p)}(0), S_m^{(p)}(1), ..., S_m^{(p)}(N_c - 1)]^T$ is the $m$th frequency domain block of the $p$th antenna, then the STBC encoded signals are [1]:

$$\begin{pmatrix} S_{2m}^{(1)}(k) & S_{2m+1}^{(1)}(k) \\ S_{2m}^{(2)}(k) & S_{2m+1}^{(2)}(k) \end{pmatrix} = \begin{pmatrix} S_{2m}(k) & -S_{2m+1}^*(k) \\ S_{2m+1}(k) & S_{2m}^*(k) \end{pmatrix}, k = 0, 1, ..., N_c-1, \quad (5)$$

where $m$ is an integer. For the case of four transmitter antennas, the space time block coding is applied to four sequential blocks as follows [15]:

$$\begin{pmatrix} S_{4m}^{(1)}(k) & S_{4m+1}^{(1)}(k) & S_{4m+2}^{(1)}(k) & S_{4m+3}^{(1)}(k) \\ S_{4m}^{(2)}(k) & S_{4m+1}^{(2)}(k) & S_{4m+2}^{(2)}(k) & S_{4m+3}^{(2)}(k) \\ S_{4m}^{(3)}(k) & S_{4m+1}^{(3)}(k) & S_{4m+2}^{(3)}(k) & S_{4m+3}^{(3)}(k) \\ S_{4m}^{(4)}(k) & S_{4m+1}^{(4)}(k) & S_{4m+2}^{(4)}(k) & S_{4m+3}^{(4)}(k) \end{pmatrix}$$
$$= \begin{pmatrix} S_{4m}(k) & -S_{4m+1}^*(k) & S_{4m+2}(k) & -S_{4m+3}^*(k) \\ S_{4m+1}(k) & S_{4m}^*(k) & S_{4m+3}(k) & S_{4m+2}^*(k) \\ S_{4m+2}(k) & -S_{4m+3}^*(k) & S_{4m}(k) & -S_{4m+1}^*(k) \\ S_{4m+3}(k) & S_{4m+2}^*(k) & S_{4m+1}(k) & S_{4m}^*(k) \end{pmatrix}. \quad (6)$$

Now $N - N_c$ zeros are added to the blocks $\mathbf{S}_m^{(p)}$ and the IFFT operation is applied to yield the time domain samples $s_m^{(p)}(n)$ ,i.e.,:

$$\begin{aligned}\mathbf{s}_m^{(p)} &= [s_m^{(p)}(0), s_m^{(p)}(1), ..., s_m^{(p)}(N-1)]^T \\ s_m^{(p)}(n) &= \frac{1}{\sqrt{N}} \sum_{k=0}^{N_c-1} S_m^{(p)}(k) e^{j\frac{2\pi kn}{N}}, n = 1, 2, ..., N.\end{aligned} \quad (7)$$

The PAPR of the $m$th frame at the $p$th antenna is defined by:

$$PAPR_m^{(p)} = \frac{\max_n \{|s_m^{(p)}(n)|^2\}}{E\{|s_m^{(p)}(n)|^2\}}, \quad (8)$$

and the ovaral PAPR at the $m$th transmission is defined by:

$$PAPR_m = \max_p PAPR_m^{(p)}. \quad (9)$$

As it can be seen, the transmitted blocks from different antennas are the conjugate or the negative conjugate of the original blocks $\mathbf{S_m}$. For example, in case of two transmitter antennas, it can be easily seen from (5) and (7) that:

$$\begin{aligned} s_{2m}^{(1)}(n) = s_{2m}(n) \quad &, \quad s_{2m}^{(2)}(n) = s_{2m+1}(n), \\ s_{2m+1}^{(1)}(n) = -s_{2m+1}^*((-n)_N) \quad &, \quad s_{2m+1}^{(2)}(n) = s_{2m}^*((-n)_N) \end{aligned} \quad (10)$$

where $(.)_N$ is mod $N$ operation. The above equations show that the time domain signals transmitted from different antennas are the conjugate or the negative conjugate of the original time domain samples $\mathbf{s_m}$. Because these operations do not change the PAPR, it is sufficient to reduce the PAPR of the original blocks $\mathbf{s_m}$ before STBC. Thus, the overall PAPR is also reduced.

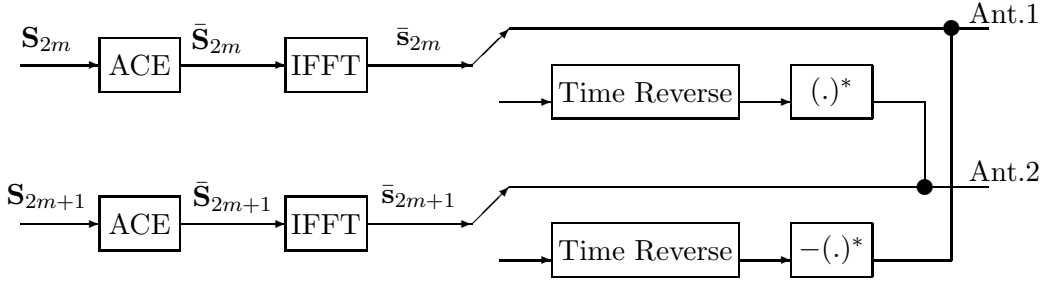

Figure 4: Block diagram of the ACE method for STBC-OFDM system with two transmitter antennas.

Fig.4 shows the block diagram of the proposed ACE method for STBC system with two transmitter antennas. As it can be seen from this figure, the iterative ACE method introduced in section 2 is applied independently to two sequential frames $\mathbf{S}_{2m}$ and $\mathbf{S}_{2m+1}$. The extended blocks $\bar{\mathbf{S}}_{2m}$ and $\bar{\mathbf{S}}_{2m+1}$ are passed through IFFT to yield low PAPR time domain vectors $\bar{\mathbf{s}}_{2m}$ and $\bar{\mathbf{s}}_{2m+1}$, respectively. Then the space time block coded signals are generated using (10) with replacing $s_{2m}(n)$ and $s_{2m+1}(n)$ by $\bar{s}_{2m}(n)$ and $\bar{s}_{2m+1}(n)$.

A similar algorithm can be proposed for STBC-OFDM system with four transmitter antennas. In this case the ACE algorithm is applied to four sequential blocks $\mathbf{S}_{4m}$, $\mathbf{S}_{4m+1}$, $\mathbf{S}_{4m+2}$ and $\mathbf{S}_{4m+3}$ to yield the extended vectors $\bar{\mathbf{S}}_{4m}$, $\bar{\mathbf{S}}_{4m+1}$, $\bar{\mathbf{S}}_{4m+2}$ and $\bar{\mathbf{S}}_{4m+3}$. Then the IFFT operation is applied to these vectors and the signal of four antennas are generated from the time domain samples $\bar{\mathbf{s}}_{4m}$, $\bar{\mathbf{s}}_{4m+1}$, $\bar{\mathbf{s}}_{4m+2}$ and $\bar{\mathbf{s}}_{4m+3}$ using time reversion and conjugate operations.

# 4 SFBC-OFDM systems, and modified ACE methods for PAPR reduction

In SFBC-OFDM systems the block $\mathbf{S}_m$ is divided to $\Gamma$ subblocks, where $\Gamma$ is the code block length. The $i$th subblock is denoted by $\mathbf{S}_{m,i}$ and is defined by:

$$\mathbf{S}_{m,i} = [S_m(i), S_m(i+\Gamma), ..., S_m(N_c - \Gamma + i)]^T, i = 0, 1, ..., \Gamma - 1. \quad (11)$$

Then the zero padded subblocks $\mathbf{X}_{m,i}$ are generated as follows:

$$\begin{aligned}\mathbf{X}_{m,i} &= [\underbrace{S_m(i), 0, \cdots, 0}_{\Gamma}, \underbrace{S_m(i+\Gamma), 0, \cdots, 0}_{\Gamma}, \cdots, \underbrace{S_m(N_c - \Gamma + i), 0, \cdots, 0}_{\Gamma}]_{N_c \times N_c} \\ &= [\underbrace{1, 0, \cdots, 0}_{\Gamma}, \underbrace{1, 0, \cdots, 0}_{\Gamma}, \cdots, \underbrace{1, 0, \cdots, 0}_{\Gamma}] \otimes \mathbf{S}_{m,i},\end{aligned} \quad (12)$$

where $\otimes$ is the kronecker product. If $Z^{-i}$ is defined as the $i$th right circular shift, then it is apparent that:

$$\mathbf{S}_m = \sum_{i=0}^{\Gamma-1} Z^{-i} \mathbf{X}_{m,i}. \quad (13)$$

In SFBC encoder the frequency domain vectors of different antennas are generated by combination of the subblocks $\mathbf{X}_{m,i}$ and their conjugates with different shifts and scaling factors. If the frequency domain vector of the $p$th antenna is denoted by $\mathbf{S}_m^{(p)}$, then it can be written as:

$$\mathbf{S}_m^{(p)} = \sum_{i=0}^{\Gamma-1} Z^{-D_i^{(p)}} \left[ a_i^{(p)} \mathbf{X}_{m,i} + b_i^{(p)} \mathbf{X}_{m,i}^* \right], p = 1, 2, ..., N_t, \quad (14)$$

where $D_i^{(p)}$ is an integer which shows the amount of shift of the $i$th subblock at the $p$th antenna and $a_i^{(p)}$ and $b_i^{(p)}$ are complex numbers. Then the frequency domain vectors $\mathbf{S}_m^{(p)}, p = 1, 2, ..., N_t$ are passed through IFFT blocks to generate the vectors $\mathbf{s}_m^{(p)}, p = 1, 2, ..., N_t$ which are the time domain samples of different antennas. If the subframes $\mathbf{X}_{m,i}, i = 0, 1, \cdots, \Gamma - 1$ are passed through IFFT block to yield the time domain subblocks $\mathbf{x}_{m,i} = [x_{m,i}(0), x_{m,i}(1), \cdots, x_{m,i}(N-1)]^T$, then it can be seen that:

$$s_m^{(p)}(n) = \frac{1}{\sqrt{\Gamma}} \sum_{i=0}^{\Gamma-1} e^{-j\frac{2\pi n D_i^{(p)}}{N}} \left[ a_i^{(p)} x_{m,i}(n) + b_i^{(p)} x_{m,i}^*((-n)_N) \right] \quad (15)$$

For example, Fig.5 shows the block diagram and frame structure of SFBC systems with two transmitter antennas and a code block length of $\Gamma = 2$. The frames $\mathbf{S}_m^{(1)}$ and $\mathbf{S}_m^{(2)}$ are generated from $\mathbf{S}_m$ as follows:

$$\begin{pmatrix} S_m^{(1)}(2\nu) & S_m^{(1)}(2\nu+1) \\ S_m^{(2)}(2\nu) & S_m^{(2)}(2\nu+1) \end{pmatrix} = \begin{pmatrix} S_m(2\nu) & S_m(2\nu+1) \\ S_m^*(2\nu+1) & -S_m^*(2\nu). \end{pmatrix}$$
$$\nu = 0, 1, ..., N_c/2 - 1 \quad (16)$$

In [4] it has been shown that if it can be assumed that the channel response at two adjacent OFDM subchannles are the same, then full diversity can be achieved. In this case the original frame $\mathbf{S}_m$ is divided by the subblocks $\mathbf{X}_{m,0}$ and $\mathbf{X}_{m,1}$ as shown bellow:

$$\mathbf{X}_{m,0} = [S_m(0), 0, S_m(2), 0, \cdots, S_m(N_c - 2), 0]^T$$
$$\mathbf{X}_{m,1} = [S_m(1), 0, S_m(3), 0, \cdots, S_m(N_c - 1), 0]^T \quad (17)$$

Then the vectors $\mathbf{S}_m^{(1)}$ and $\mathbf{S}_m^{(2)}$ can be written in the form of (14), i.e.,:

$$\mathbf{S}_m^{(1)} = \mathbf{X}_{m,0} + Z^{-1}\mathbf{X}_{m,1},$$

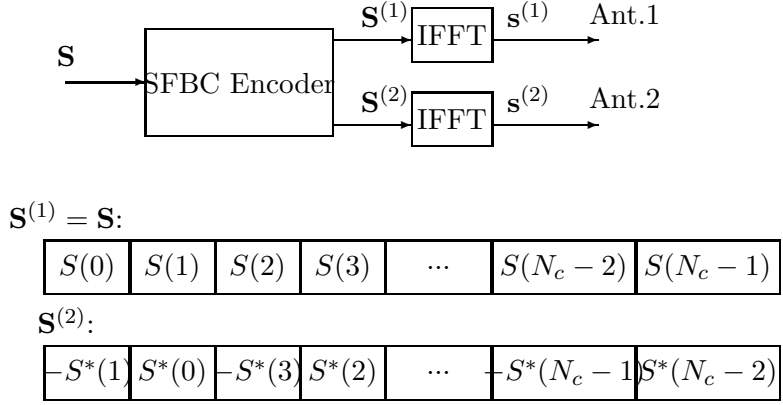

$\mathbf{S}^{(1)} = \mathbf{S}$:

| $S(0)$ | $S(1)$ | $S(2)$ | $S(3)$ | $\cdots$ | $S(N_c-2)$ | $S(N_c-1)$ |

$\mathbf{S}^{(2)}$:

| $-S^*(1)$ | $S^*(0)$ | $-S^*(3)$ | $S^*(2)$ | $\cdots$ | $-S^*(N_c-1)$ | $S^*(N_c-2)$ |

Figure 5: Block diagram and frame structure of SFBC-OFDM system with two transmitter antennas.

$$\mathbf{S}_m^{(2)} = -Z^{-1}\mathbf{X}_{m,0}^* + \mathbf{X}_{m,1}^* \qquad (18)$$

In case of $N_t = \Gamma = 4$, the encoding procedure is as follows:

$$\begin{pmatrix} S_m^{(1)}(4\nu) & S_m^{(1)}(4\nu+1) & S_m^{(1)}(4\nu+2) & S_m^{(1)}(4\nu+3) \\ S_m^{(2)}(4\nu) & S_m^{(2)}(4\nu+1) & S_m^{(2)}(4\nu+2) & S_m^{(2)}(4\nu+3) \\ S_m^{(3)}(4\nu) & S_m^{(3)}(4\nu+1) & S_m^{(3)}(4\nu+2) & S_m^{(3)}(4\nu+3) \\ S_m^{(4)}(4\nu) & S_m^{(4)}(4\nu+1) & S_m^{(4)}(4\nu+2) & S_m^{(4)}(4\nu+3) \end{pmatrix}$$
$$= \begin{pmatrix} S_m(4\nu) & S_m(4\nu+1) & S_m(4\nu+2) & S_m(4\nu+3) \\ -S_m^*(4\nu+1) & S_m^*(4\nu) & -S_m^*(4\nu+3) & S_m^*(4\nu+2) \\ S_m(4\nu+2) & S_m(4\nu+3) & S_m(4\nu) & S_m(4\nu+1) \\ -S_m^*(4\nu+3) & S_m^*(4\nu+2) & -S_m^*(4\nu+1) & S_m^*(4\nu) \end{pmatrix}. \qquad (19)$$

In this case, if $\mathbf{X}_{m,i}, i = 0, 1, 2, 3$ are defined based on (12), then it can be written as:

$$\mathbf{S}_m^{(1)} = \mathbf{X}_{m,0} + Z^{-1}\mathbf{X}_{m,1} + Z^{-2}\mathbf{X}_{m,2} + Z^{-3}\mathbf{X}_{m,3}$$

$$\mathbf{S}_m^{(2)} = -Z^{-1}\mathbf{X}_{m,0}^* + \mathbf{X}_{m,1}^* - Z^{-3}\mathbf{X}_{m,2}^* + Z^{-2}\mathbf{X}_{m,3}^*$$

$$\mathbf{S}_m^{(3)} = Z^{-2}\mathbf{X}_{m,0} + Z^{-3}\mathbf{X}_{m,1} + \mathbf{X}_{m,2} + Z^{-1}\mathbf{X}_{m,3}$$

$$\mathbf{S}_m^{(4)} = -Z^{-3}\mathbf{X}_{m,0}^* + Z^{-2}\mathbf{X}_{m,1}^* - Z^{-1}\mathbf{X}_{m,2}^* + \mathbf{X}_{m,3}^* \quad (20)$$

Similar to (8) and (9), the PAPR of different antennas is defined and the overall PAPR is the maximum PAPR among antennas. To reduce the PAPR, the symbols $\mathbf{S}_m$ are extended to $\mathbf{S}_m + \mathbf{C}_m$. If the extension vector $\mathbf{C}_m$ is divided by subblocks $\mathbf{C}_{m,i}$ similar to (12), the time domain signal of the $p$th antenna becomes:

$$\bar{s}_m^{(p)}(n) = s_m^{(p)}(n) + \frac{1}{\sqrt{N}} \sum_{i=0}^{\Gamma-1} \sum_{k=0}^{N_c-1} e^{-j\frac{2\pi n D_i^{(p)}}{N}} \left[ a_i^{(p)} C_{m,i}(k) + b_i^{(p)} C_{m,i}^*(k) \right] e^{-j\frac{2\pi nk}{N}} \quad (21)$$

The main goal in SFBC case is the determination of the extension vector $\mathbf{C}_m$ such that the maximum PAPR among the antennas is minimized. Thus, to minimize the maximum PAPR of SFBC system, the following optimization problem must be solved:

$$\min_{\mathbf{C}_m} \left\{ \max_{n,p} \left\{ s_m^{(p)}(n) + \frac{1}{\sqrt{N}} \sum_{i=0}^{\Gamma-1} \sum_{k=0}^{N_c-1} e^{-j\frac{2\pi n D_i^{(p)}}{N}} \left[ a_i^{(p)} C_{m,i}(k) + b_i^{(p)} C_{m,i}^*(k) \right] e^{-j\frac{2\pi nk}{N}} \right\} \right\}$$

$$\text{where} \quad \mathbf{C_m} = \sum_{i=0}^{\Gamma-1} Z^{-i} \mathbf{C}_{m,i} \quad Subject \quad to \quad \|\mathbf{C}_m\|^2 \leq \Delta P \quad (22)$$

To find the suboptimum solution of (22), in the following subsections two methods have been proposed.

### 4.1 Sub-ACE method

From (15) it can be seen that the signals of $N_t$ antennas are the combination of the time domain samples of the subblocks $x_{m,i}(n)$ and their conjugates. Thus if the PAPR of these subframes are reduced, the PAPR at all the antennas is also reduced. It means that instead of solving (22), the optimization problem is solved as follows:

$$\min_{\mathbf{C}_{m,i}} \left\{ \max_n \{x_{m,i}(n) + \frac{1}{\sqrt{N/\Gamma}} \sum_{k=0}^{N_c-1} C_{m,i}(k)e^{-j\frac{2\pi nk}{N}} \} \right\}, i = 0, 1, \cdots, \Gamma - 1 \quad (23)$$

Then the time domain subframes $\mathbf{x}_{m,i}$ are modified as:

$$\bar{x}_{m,i}(n) = x_{m,i} + \frac{1}{\sqrt{N/\Gamma}} \sum_{k=0}^{N_c-1} C_{m,i}(k)e^{-j\frac{2\pi nk}{N}}. \quad (24)$$

Afterward the signal of $N_t$ antennas are generated based on (15) with replacing $x_{m,i}(n)$ by $\bar{x}_{m,i}(n)$. This method is based on the PAPR reduction of the subframes, thus the name Sub-ACE method. Fig.6 shows the block diagram of the Sub-ACE method. A similar method has been proposed in [16] but the PAPR of the subframes is reduced by clipping and filtering instead of the ACE method.

Note that in the Sub-ACE method, the PAPRs of the subframes are reduced independently. When these subframes are recombined, it is possible to make new peaks because the phase of the subframes may be equal in some time samples. It is clear that when the number of subblocks grows (which is equal to the code length), the peak generation in the composition stage

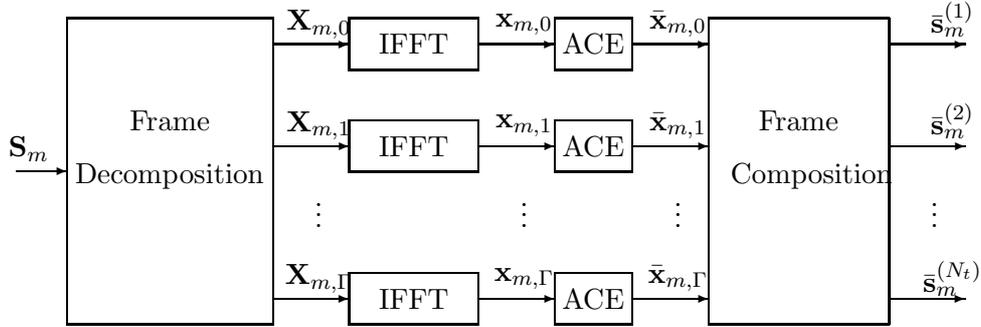

Figure 6: Block diagram of the Sub-ACE method for PAPR reduction of SFBC-OFDM systems with $N_t$ transmitter antennas with a code block length of $\Gamma$.

is more probable.

## 4.2 Selective-ACE method

Another approach to find the suboptimum solution of (22) is the application of ACE method to the antenna signal with maximum PAPR and construction of the other antenna signals from this signal. Fig.7 shows the block diagram of this method. This figure show that the frequency domain vectors of different antennas are generated from original frame $\mathbf{S}_m$ using (12) and (14). Then the time domain samples are derived using IFFT operation. Among these signals, the one with the maximum PAPR is selected which is denoted by $\mathbf{S}_m^{(q)}$. The time domain signal of selected antenna is passed through the clipping and FFT operations. Then $N - N_c$ out-of-band components are removed and the in-band components $S_m^{(q)}(k), k = 0, 1, \cdots, N_c - 1$ are mapped to the allowable regions shown in Fig.1 to prevent decreasing the distance between the constellation points. The resulting vector is $\bar{\mathbf{S}}_m^{(q)}$. Now the signals of the other antennas must be generated from these extended symbols. For this aim, the modified frame $\bar{\mathbf{S}}_m$ is regenerated from $\bar{\mathbf{S}}_m^{(q)}$ and the signal of the other antennas are produced from the modified symbols $\bar{S}_m(k)$. Note that in case of the two transmitter antennas described in (16) the frame $\bar{\mathbf{S}}_m$ can be generated from $\bar{\mathbf{S}}_m^{(1)}$ or $\bar{\mathbf{S}}_m^{(2)}$ as follows:

$$\bar{S}_m(k) = \bar{S}^{(1)}(k), k = 0, 1, ..., N_c - 1,$$

$$\bar{S}_m(k) = \begin{cases} -\bar{S}_m^{(2)*}(2\nu + 1) & \text{if } k = 2\nu \\ \bar{S}_m^{(2)*}(2\nu) & \text{if } k = 2\nu + 1 \end{cases}, \quad (25)$$

Similarly from (19), a relationship can be derived to generate $\bar{\mathbf{S}}_m$ from $\bar{\mathbf{S}}_m^{(q)}(q = 1, 2, 3, 4)$ in case of the four transmitter antennas.

The procedure of selecting the antenna with the maximum PAPR, applying clipping, filtering, mapping and constructing the signal of other antennas is performed iteratively. This method is called Selective-ACE because at each iteration, the ACE method is applied to the selected antenna with the maximum PAPR.

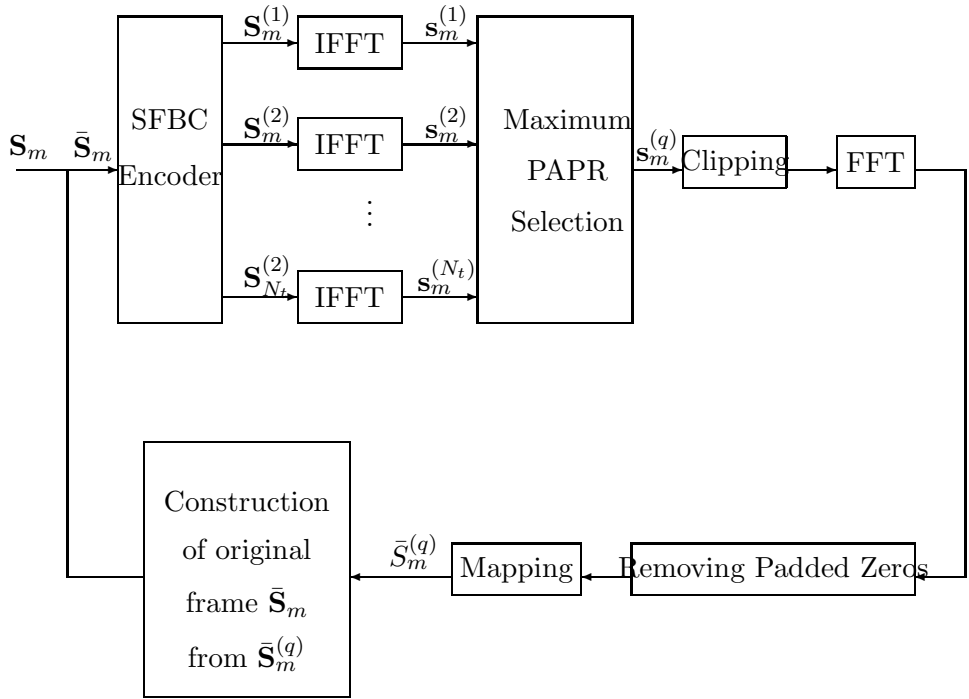

Figure 7: Block diagram of the Selective-ACE method for PAPR reduction of SFBC-OFDM systems with $N_t$ transmitter antennas.

## 5 Computational Complexity

In this section, the computational complexity of the proposed methods is discussed. The original ACE method and the proposed ACE algorithms for STBC and SFBC cases include five steps shown in Table.1. This table summarizes the order of complexity at each step of the algorithms. Note that the complexity of the ACE method in STBC case is the same as the complexity of the ACE method in the single antenna case. As it can be seen from this table, the complexity of the Sub-ACE method is even less than the original ACE method. Because one $N$ points IFFT operation in the original ACE method is converted to $N_t$ IFFTs with the size of $N/N_t$ points. The complexity of the Selective-ACE is higher than the original ACE method because at each iteration, $N_t$ IFFTs of length $N$ must be calculated. At the step of finding the antenna with the maximum PAPR, the amplitude of the $NN_t$ time domain samples must be calculated and the maximum value must be found.

|  | ACE [14] | Sub-ACE | Selective ACE |
|---|---|---|---|
| IFFT | $N \log N$ | $N \log \frac{N}{N_t}$ | $N_t N \log N$ |
| Amlitude Calculation | $N$ | $N$ | $NN_t$ |
| Amlitude comparison with threshold | $N$ | $N$ | $NN_t$ |
| FFT | $N \log N$ | $N \log \frac{N}{N_t}$ | $N \log N$ |
| Mapping | $N_c$ | $N_c$ | $N_c$ |

Table 1: The order of the complexities of the five steps of the original and the modified ACE methods for the case of the single antenna and the SFBC-OFDM systems.

## 6 Simulation results

In our simulations, the OFDM frames with $N_c = 256$ subcarriers have been assumed with QPSK modulation. To find the peak values and also to estimate the PAPR of the analogue signal, the oversampling ratio of 4 has been used. The two cases of STBC-OFDM and SFBC-OFDM have been simulated. In the calculation of the PAPR the ratio of the maximum power to the initial power (before constellation extension) has been considered, thus the power increase of the PAPR reduction algorithm has been taken into account. The performance of the proposed methods is evaluated by the Complementary Cumulative Density Function (CCDF) of the PAPR which is defined as:

$$CCDF(PAPR_0) = Pr\{PAPR \geq PAPR_0\}. \tag{26}$$

Fig.8 shows the performance of the ACE method in the STBC-OFDM system with the two transmitter antennas using various numbers of iteration. The optimum clipping level in this case is $4.86dB$ above the average power of the OFDM frames. As it was mentioned before, the performance of the ACE method in STBC-OFDM is similar to the single antenna OFDM systems and it does not depend on the number of transmitter antennas. It can be seen from this figure that the PAPR reduction at the probability of $10^{-4}$ is about $1.2dB$, $2.6dB$, $3.5dB$ and $3.7dB$ after one, three, five and seven iterations, respectively.

Fig.9 shows the performance of the modified ACE methods in the SFBC-OFDM system with two transmitter antennas and QPSK modulation. In this case, two proposed methods, Sub-ACE and Selective-ACE algorithms, have been simulated. In the Selective-ACE method, the optimum clipping level is $4.86dB$ above the average power of OFDM frames and in the Sub-ACE algorithm is $4.86dB$ above the average power of the sub-frames. It is apparent that the performance of the Selective-ACE method is better than the Sub-ACE method. The main reason is the peak regrowth in the step of subframe combination in the Sub-ACE algorithm.

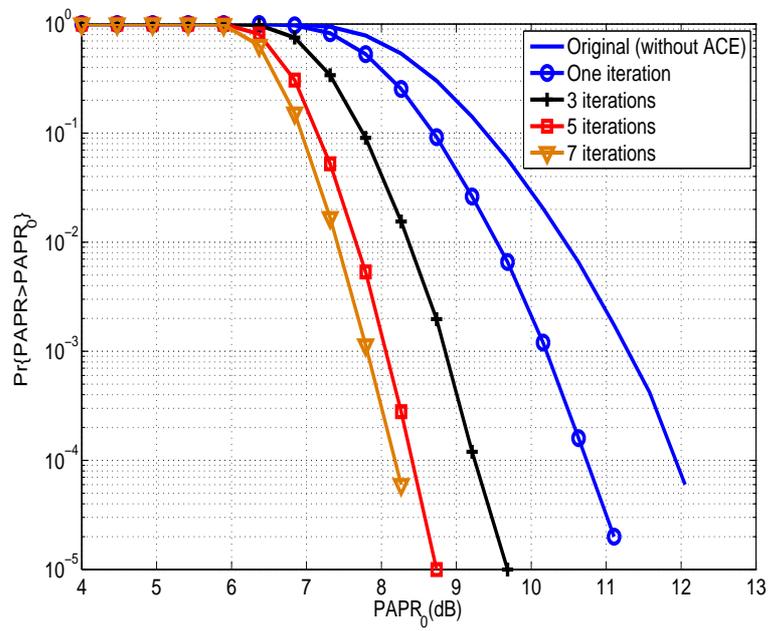

Fig. 8: The performance of the ACE method for STBC with 2 transmitters after one, 3, 5 and 7 iterations.

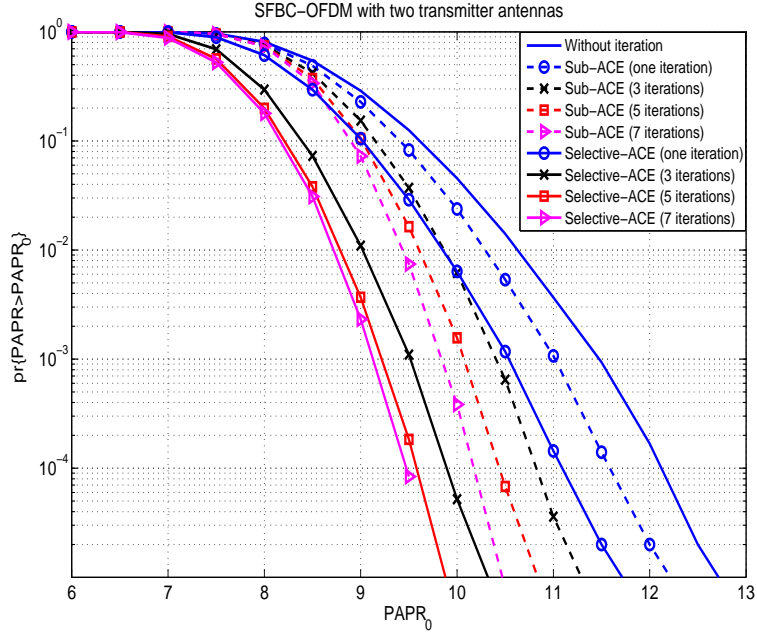

Fig. 9: The performance of the ACE method for the SFBC-OFDM with 2 transmitters after one, 3, 5 and 7 iterations.

Fig.10 shows the CCDF of the PAPR in case of the SFBC-OFDM system with four transmitter antennas based on the encoding structure of (19). It is apparent from this figure that the performance of the Sub-ACE method is worse than the two antenna case. As it was mentioned before when the number of subblocks increases the probability of peak regeneration in subframe composition step also increases. The Selective-ACE method has a good performance in this case. To compare the results for two and four antenna cases in Fig.11, the PAPR at probability of $10^{-4}$ has been plotted versus the number of iterations. Since the performance of the ACE method in the STBC is similar to the performance of the ACE in the single antenna OFDM system, it is a good upper bound on the performance of the other methods. As it can be seen from Fig.11, the Selective-ACE method has only about $0.7dB$ degradation in comparison to the ACE in the STBC case at the 7th iteration, while the performance degradations are $1.7dB$ and $3.1dB$ for the Sub-ACE method in case of the two and four transmitter antennas, respectively.

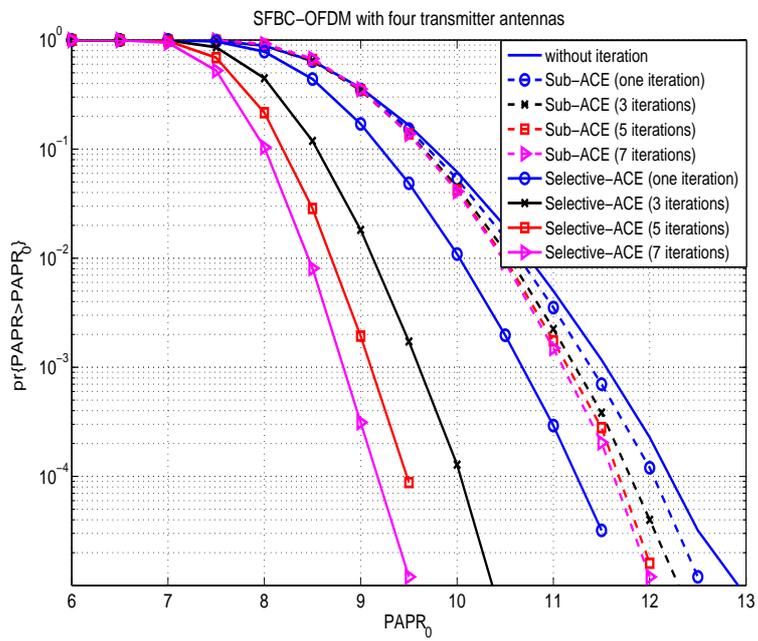

Fig. 10: The performance of the ACE method for the SFBC-OFDM with 4 transmitters after one, 3, 5 and 7 iterations.

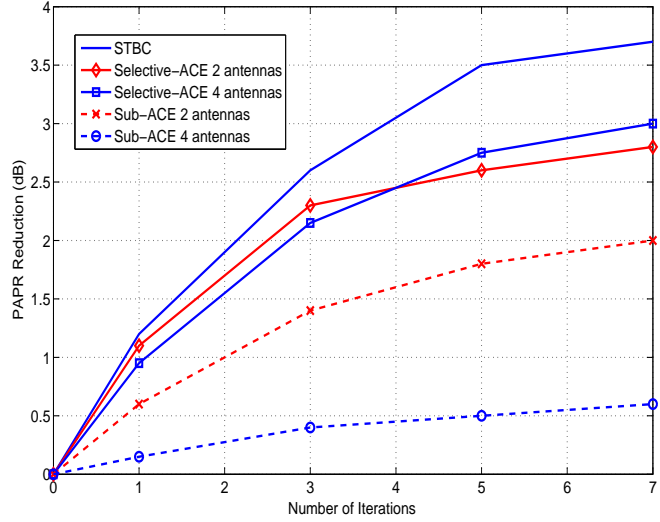

Fig. 11: The PAPR reduction versus the number of iterations for the STBC and the SFBC systems with two and four transmitter antennas.

# 7 Conclusion

In this paper the extension of the ACE for PAPR reduction for the space time and space frequency block coded OFDM systems has been discussed. It is shown that the ACE method can be used in the STBC case as an independent block before the spatial encoding and its performance is similar to the ACE in single antenna OFDM systems. In the SFBC case, we have proposed two new algorithms: one is based on applying the ACE on the subframes (Sub-ACE) and the other one is based on applying the ACE on the antenna with the maximum PAPR (Selective-ACE). Simulation results show that both algorithms converge but the performance of the Sub-ACE method degrades due to peak regrowth in the step of subframe recombination. This degradation is more obvious when the code block length increases. The performance of the Selective-ACE is better and is close to the performance of the ACE method in single antenna OFDM systems. This improvement is more pronounced when the number of transmission antennas increases. In the Selective-ACE method the PAPR reduction of about $3dB$ can be achieved in both cases of the two and the four transmitter antennas.